\newcommand{\vtr}[1]{\boldsymbol{#1}}

\documentclass[
reprint,
superscriptaddress, amsmath,amssymb,aps,prl
]{revtex4-2}

\usepackage{graphicx}
\usepackage{dcolumn}
\usepackage{bm}
\usepackage{chemformula}
\usepackage{multirow}
\usepackage{array}
\usepackage{verbatim}
\usepackage[normalem]{ulem} 
\usepackage{hyperref}
\hypersetup{colorlinks=true,citecolor=blue,linkcolor=blue,urlcolor  = blue, breaklinks=true}
\usepackage{upgreek}
\usepackage{xcolor}
\usepackage{ulem}
\newcolumntype{P}[1]{>{\centering\arraybackslash}p{#1}}

\begin{document}


\title{Field-like Perturbation Enabled Six-state Readout in Triaxial $\alpha$-\ch{Fe2O3}$\vert$\ch{Pt} Bilayers}


\author{Aditya A. Wagh}
\email[]{adityawagh@iisc.ac.in}
\affiliation{%
 Department of Physics, Indian Institute of Science, Bengaluru, INDIA 
}%

\author{Shwetha G. Bhat}
\email[]{shwetha@iisc.ac.in}
\affiliation{%
Department of Physics, Indian Institute of Science, Bengaluru, INDIA 
}%

\author{Krishna Jha}
\affiliation{%
Department of Physics, Indian Institute of Science, Bengaluru, INDIA 
}%

\author{Aiswarya Sukumaran}
\affiliation{%
Department of Physics, Indian Institute of Science Education and Research, Tirupati, INDIA 
}%

\author{P. S. Anil Kumar}
\email[]{anil@iisc.ac.in}
\affiliation{%
Department of Physics, Indian Institute of Science, Bengaluru, INDIA 
}%

\begin{abstract}
Canted antiferromagnet $\alpha$-\ch{Fe2O3}$\vert$\ch{Pt} bilayers offer a platform for multistate memory, but spin Hall magnetoresistance (SMR) readout cannot distinguish states separated by $180^{\circ}$, limiting detection to three states. We show that a field-like perturbation lifts the degeneracy of opposite states in $\alpha$-\ch{Fe2O3}, enabling the resolution of six states in the SMR signal. We introduce a dual-modulation SMR method to mitigate thermal drifts and enable reliable readout. Computations of first- and second-harmonic SMR reveal the role of exchange, anisotropy and field-like perturbations in resolving the states.
\end{abstract}

\maketitle

\section{\label{sec:level1}INTRODUCTION}

Manipulation of the N\'{e}el vector ($\vtr{n}$) in antiferromagnets (AFM) via spin–orbit torque (SOT) offers a robust pathway for electrical control of AFM order.  The associated dynamics occur in the ultra-fast regime and are resilient to external perturbations, far exceeding conventional ferromagnetic limits \cite{Zarzuela2017, Lebrun2020, Yang2024, kanj2023, Baltz2018}. Recently, non-collinear AFMs \cite{Zhu2024,Lee2025,Raju2025,Goren2025,Gao2025,Ogawa2026,Zhou2026,Rimmler2025,Romaguera2024,Hoogeboom2021,Wagh2022,Wagh2023,Agarwal2024,Han2020,Aqeel2015,Aqeel2016} have emerged as an important class of materials hosting phenomena such as nonreciprocal spin transport \cite{Guckelhorn2023,kanj2023}, altermagnetism \cite{Hoyer2025} and unconventional spin Hall magnetoresistance (SMR) \cite{Uchimura2025}. The interfacial spin-current framework enables current-induced switching of $\vtr{n}$ in AFM/heavy metal bilayers, such as \ch{NiO}$\vert$\ch{Pt} \cite{Chen2018, Moriyama2018, Baldrati2019}, $\alpha$-\ch{Fe2O3}$\vert$\ch{Pt} \cite{Zhang2019, Cheng2020, Zhang2022, Cogulu2022} and facilitates sensitive detection of the orientation of $\vtr{n}$ via probing SMR, arising from interfacial spin-current absorption \cite{Althammer2013, Chen2013,Lebrun2019}.

In general, due to SMR's invariance under the two-fold rotation group $C_{2z}$, the SMR signal ($\propto\langle m_{x} m_{y} \rangle$) cannot distinguish between diametrically opposite states related by a $180^{\circ}$ rotation. Within the constraints imposed by SMR-probing, the tri-state memory demonstration in $\alpha$-\ch{Fe2O3} \cite{Cheng2020} resolved three orientations of $\vtr{n}$ rather than all six symmetry-allowed orientations. In this study, we show that a field-like (FL) perturbation, induced via a magnetic field or a current-induced torque, lifts the degeneracy in SMR amplitudes pertaining to opposite $\vtr{n}$-orientations in canted AFMs, enabling unambiguous resolution of all six states in $\alpha$-\ch{Fe2O3}$\vert$\ch{Pt} bilayers. With our magnetization dynamics model, we numerically compute first (1$\omega$) and second harmonic (2$\omega$) responses of the SMR signal to FL perturbations to the system at equilibrium governed by the in-plane triaxial anisotropy.  Experimentally, the application of perturbative field to each of the six magnetically written states gave unique signal for each state, thus resolving all of them. A detailed numerical analysis of 2$\omega$ SMR gave crucial insights into the interplay of the underlying interactions and revealed suitable parameters for the SMR resolution. We theoretically put forward a two-step current-only protocol for a six-state memory readout. Additionally, to mitigate thermal drift in SMR signals, we propose dual-modulation (current and field) SMR detection and demonstrate a temperature-robust, reliable state readout.

\section{Tri-state vs. Six-state memory}
\label{Six-state}

\begin{figure}
\includegraphics[scale=0.5]{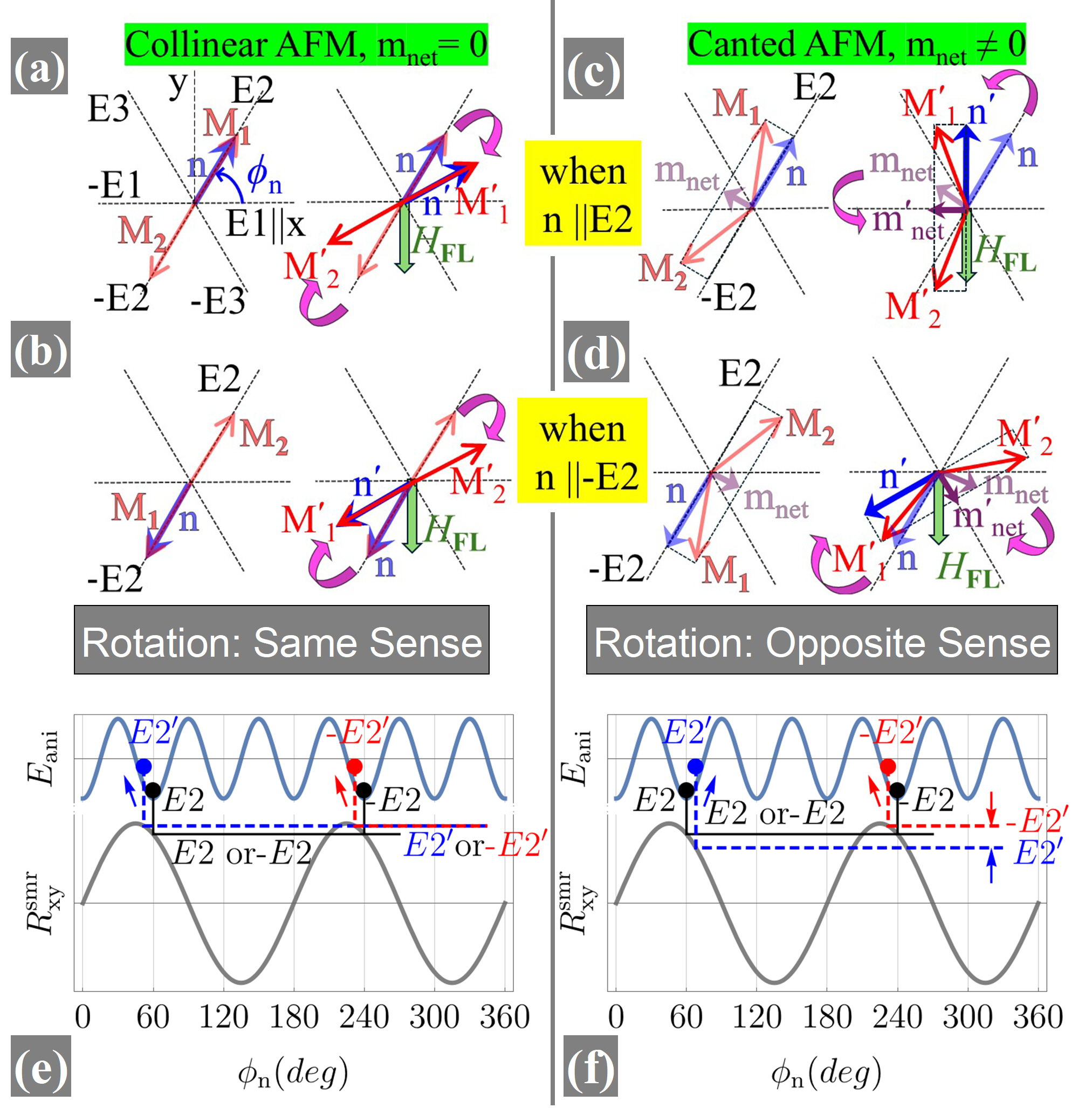}
\caption{\label{fig:1}Response of spin configurations and SMR to applied field-like torque (due to $\vtr{H}_{\text{FL}}$) in collinear and canted AFMs with tri-axial anisotropy. (a), (b) Initial (\textbf{E2} or -\textbf{E2}) and final (\textbf{E2$^\prime$} or \textbf{–E2$^\prime$}) states in collinear AFMs; (c), (d) corresponding states in canted AFMs. (e), (f) Angular dependence of $E_{\text{ani}}$ and $R^{\text{SMR}}_{\text{xy}}$ with $\phi_{\text{n}}$ for collinear and canted AFMs, respectively, where $\phi_{\text{n}}$ is measured from $\hat{x}$. Opposite rotation in canted AFMs lifts the degeneracy between \textbf{E2} and \textbf{–E2}, yielding distinct SMR for \textbf{E2$^\prime$} and \textbf{–E2$^\prime$}, unlike collinear AFMs.}
\end{figure}

In collinear AFMs, the sublattice magnetizations $\vtr{M}_{\text{1}}$ and $\vtr{M}_{\text{2}}$ align antiparallel due to exchange interaction ($J$), yielding zero net moment $\vtr{m}_{\text{net}}=\frac{\vtr{M}_{\text{1}}+\vtr{M}_{\text{2}}}{2}$, and $\vtr{n} = \frac{\vtr{M}_{\text{1}}-\vtr{M}_{\text{2}}}{2}$. In canted AFMs, Dzyaloshinskii–Moriya (DM) interaction \cite{Moriya1960} induces finite canting, generating a small $\vtr{m}_{\text{net}}$. Magnetocrystalline anisotropy ($K_{\text{ani}}$) further stabilizes $\vtr{M}_{\text{1,2}}$. In systems such as $\alpha$-\ch{Fe2O3} and \ch{NiO}, tri-axial anisotropy in the (0001) and (111) planes, respectively \cite{Cheng2020, Lee2012} can be described by the anisotropy energy $E_{\text{ani}}\propto -\cos{6\phi_{\text{n}}}$, where $\phi_{\text{n}}$ is the angle of $\vtr{n}$ measured from the $\hat{x}$-axis (Fig. \ref{fig:1}). This yields six stable states (\textbf{E1}, \textbf{E2}, \textbf{E3}, \textbf{-E1}, \textbf{-E2}, \textbf{-E3}) separated by 60$^\circ$.  While SMR resolves \textbf{E1}, \textbf{E2}, and \textbf{E3}, the pair of opposite states (\textbf{E\textit{i}} from \textbf{-E\textit{i}}, \textbf{\textit{i}} = 1, 2 and 3) remains indistinguishable due to identical amplitudes. To overcome this limitation, we examine the response of collinear and canted AFMs to a small field-like torque (FLT) under single-domain initialization. In collinear AFMs, FLT on opposite states (\textbf{E} or \textbf{-E}), yields same sense of rotation of $\vtr{n}$ (Fig. \ref{fig:1}(a)-(b)). In contrast, canted AFMs exhibit opposite sense of rotation (Fig. \ref{fig:1}(c)-(d)), lifting the degeneracy in SMR to produce distinct response for \textbf{E} and \textbf{-E} (see Fig. \ref{fig:1}(f) and Section \ref{1w}), unlike the indistinguishable states in collinear systems (Fig. \ref{fig:1}(e)). Consequently, all six states can be resolved unambiguously, as demonstrated experimentally in $\alpha$-\ch{Fe2O3} in  Section \ref{results}, establishing an FLT-based route for six-state readout beyond conventional tri-state SMR detection.

\section{Results and Discussion}\label{results}
\subsection{SMR-response to static field-like perturbation}\label{1w}
\begin{figure}
\includegraphics[width=0.5\textwidth,keepaspectratio]{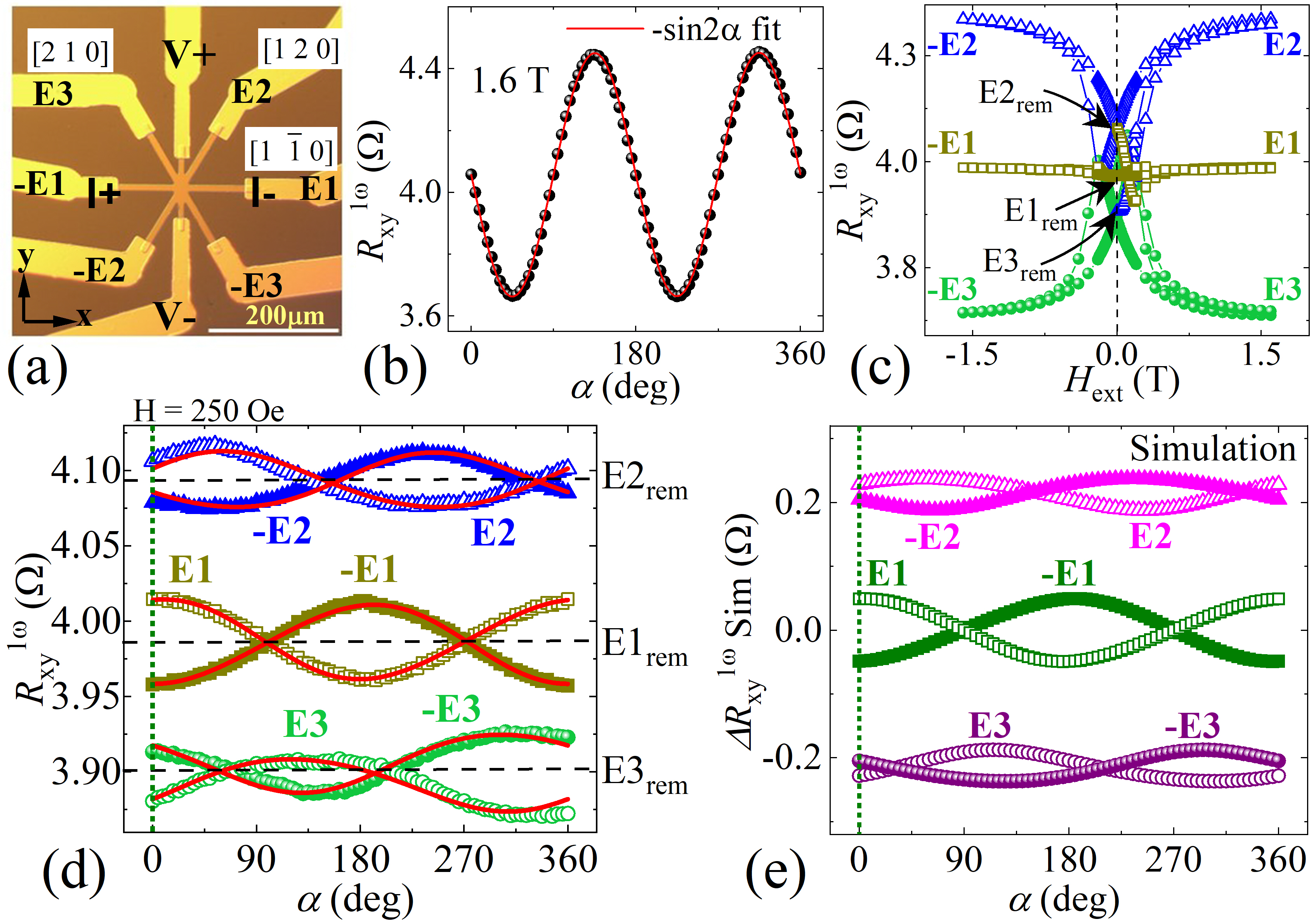}
\caption{\label{fig:2}(a) Optical micrograph of the eight-terminal $\alpha$-\ch{Fe2O3} device. (b) Field-variation of $R_{\text{xy}}^{\,\text{1}\omega}$ for $\vtr{H}_{\text{ext}}\perp\vtr{n}$ and $\vtr{H}_{\text{ext}} \parallel \vtr{m}_{\text{net}}$ for each state \textbf{E}. (c) The single-domain response of $\sin{2\alpha}$ in $R_{\text{xy}}^{\,\text{1}\omega}$ at $H_{\text{ext}}$ = 1.6 T and $I_{\text{rms}}$ of 2 mA. (d) $R_{\text{xy}}^{\,\text{1}\omega}$ vs. $\alpha$ scans for the six states at $H_{\text{ext}}$ = 250 Oe and $I_{\text{rms}} = $ 2 mA along with the data fits. (e) Simulated $R_{\text{xy}}^{\,\text{1}\omega}$ vs. $\alpha$ curves for the six states.
}
\end{figure}

$\alpha$-\ch{Fe2O3}(17.6 nm)$\vert$\ch{Pt}(2.3 nm) film was grown using pulsed laser deposition (refer \cite{supplemental}) and patterned into an eight-terminal device (Fig. \ref{fig:2}(a)) with bars along \textbf{E1}, \textbf{E2}, and \textbf{E3} (180 $\mu m \times$ 10 $\mu m$). The current ($I_\textbf{rms}$ = 2 mA, 333 Hz) is passed using Keithley 6221 DC/AC current source along \textbf{E1} bar (Fig. \ref{fig:2}(a)). The transverse SMR signal ($R_{\text{xy}}^{\,\text{1}\omega}=V_{\text{xy}}^{\,\text{1}\omega}/I_{\textbf{rms}}$) was measured using a lock-in amplifier (Stanford Research System SR830). In-plane angular ($\alpha$-scan, defined relative to $\hat{x}$) and field-dependent measurements were performed at 305 K using a LakeShore electromagnet with a custom-made sample rotation stage for the fields $H_{\text{ext}}\leq$ 1.6 T. High-field measurements were done using a Physical Property Measurement System (PPMS), Quantum Design.

At 1.6 T, $R_{\text{xy}}^{\,\text{1}\omega}$ follows a clear $-\sin{2\alpha}$-dependence (Fig. \ref{fig:2}(b)) indicating formation of a single-domain and its free rotation with field ($H_{\text{ext}} \gg H_{\text{ani}}$, the anisotropy field) \cite{Zhang2019, Cheng2020, Cogulu2022}. $H_{\text{ext}}$ scans were performed up to $\pm$1.6 T for each state (\textbf{E}) with $\vtr{H}_{\text{ext}}\perp\vtr{n}$ and $\vtr{H}_{\text{ext}} \parallel \vtr{m}_{\text{net}}$, and the corresponding $1\omega$ signals $R_{\text{xy}}^{\,\text{1}\omega}$ are shown in Fig. \ref{fig:2}(c). At $H_{\text{ext}}=\pm$1.6 T, three pairs of states (\textbf{E1}/\textbf{-E1}, \textbf{E2}/\textbf{-E2}, \textbf{E3}/\textbf{-E3}) are magnetically written and upon subsequent removal of $H_{\text{ext}}$, they attain remanent $R_{\text{xy}}^{\,\text{1}\omega}$ values ($E1_{\text{rem}}$, $E2_{\text{rem}}$ and $E3_{\text{rem}}$). The virgin curves starting at $H_{\text{ext}}$ = 0 reflect a history of previous state written. Thus, each of the six single-domain states can be written using the appropriate application of $H_{\text{ext}}=$ 1.6 T ($\vtr{H}_{\text{ext}}\perp\vtr{n}$ and $\vtr{H}_{\text{ext}} \parallel \vtr{m}_{\text{net}}$). Next, we introduced static FL perturbation ($H_{\text{ext}} = $ 250 Oe) to the written state and study its SMR response by recording the $\alpha$-scan. Measurements (Fig. \ref{fig:2}(d)) reveal 360$^\circ$ periodic $R_{\text{xy}}^{\,\text{1}\omega}$ signals for all six states. We fitted the SMR curves by including a common ordinary Hall effect contribution (due to sample misalignment \cite{Wagh2022}). In particular, six sinusoidal signals are observed with successive 60$^\circ$ phase shifts. The SMR signals pertaining to oppositely oriented states modulate out of phase in $\alpha$, lifting the degeneracy of SMR-values and enabling their identification. Notably, at $\alpha$ = 0, distinct SMR values help to resolve all six states. Similar experiments on a second device with Hall-bar geometry (Fig. S2 \cite{supplemental}) showed consistent results. 

Understanding the underlying mechanism behind the 360$^\circ$ periodic $R_{\text{xy}}^{\,\text{1}\omega}$ signals observed for the six states is crucial for the readout. Hence, to capture the dynamic response of sublattice magnetization, we numerically solved the torque balance equation of the system as discussed in the appendix \ref{LLG}. In Eq. \ref{eq:HeffLLG}, we used parameter values of $J$ = 930 T, DM interaction $|\vtr{D}|$ = 2.5 T, easy-plane anisotropy ${K}_{\text{plane}}$ = 0.17 T, tri-axial anisotropy ${K}_{\text{ani}}$= 1$\times 10^{-5}$ T with Gilbert damping parameter $\alpha_{\text{damp}}$ = 5$\times 10^{-4}$ and gyromagnetic ratio $\gamma$ = 28.0113 GHz/T, consistent with the literature \cite{Sulymenko2017, Cogulu2022, Yang2024, Sheng2025}. The equilibrium spin-configuration $\vtr{M}_{1,2}$ was used to calculate the transverse SMR response ($R_{\text{xy}} \propto \langle m_{(1,2)\text{x}} \cdot m_{(1,2)\text{y}}\rangle$). Further, the angular-dependent first harmonic SMR ($R_{\text{xy}}^{\,\text{1}\omega}$) at various $H_{\text{ext}}$ values was simulated. For $H_{\text{ext}} \gg H_{\text{ani}}$, the simulated curve $R_{\text{xy}}^{\,\text{1}\omega}$ exhibits an expected $-\sin{2\alpha}$ behavior in $\alpha$ with 180$^\circ$ periodicity (Fig. S3 \cite{supplemental}). After initialization of a single domain of choice (state $\pm$\textbf{E\textit{i}}, \textbf{\textit{i}} = 1, 2 and 3) under the perturbation $H_{\text{ext}} \ll H_{\text{ani}}$, simulations yielded a clear 360$^\circ$-periodic $R_{\text{xy}}^{\,\text{1}\omega}$, with a 180$^\circ$ phase-shift between opposite states (Fig. \ref{fig:2}(e)), consistent with the experimental observations. Baseline shifts observed experimentally (\textbf{E1} to \textbf{E3}), arising from thermal effects, are discussed in Section \ref{sideband}. Our model shows that spontaneous canting enables $\vtr{m_{\text{net}}}$ to respond to $H_{\text{ext}}$, producing state-dependent torques that help distinguish between \textbf{E} and \textbf{-E} states (Fig.~\ref{fig:1}(c),(d),(f)).
In contrast, the SMR response is indistinguishable for \textbf{E} and \textbf{-E} states in collinear AFMs (Fig.~\ref{fig:1}(a),(b),(e)) and our numerical simulations reveal a 180$^\circ$-periodic $R_{\text{xy}} \propto\sin{(2\alpha+\text{phase})}$ (Fig. S4 \cite{supplemental}).

Solving for an analytical form of Eq. \ref{eq:HeffLLG} is intricate due to the coupled non-linear differential nature of the equations. However, after the initialization of a single domain of choice, under subsequent in-plane $H_{\text{ext}} \ll H_{\text{ani}}$ ($\vtr{H_{\text{ext}}} = H_{\mathrm{ext}} (\cos\phi_{\text{H}}\,\hat{x}+\sin\phi_{\text{H}}\,\hat{y})$, $\phi_{\text{H}}$ is measured from $\hat{x}$), the system can be described by a simplified effective Hamiltonian comprising Zeeman and in-plane triaxial anisotropy terms as, $E_{\text{total}} = \left| \vtr{n}\cdot\vtr{H_{\text{ext}}} \right|^{2} - \left(\vtr{m}_{\text{net}}\cdot\vtr{H_{\text{ext}}}\right) -{K}_{\text{ani}} \cos6 \phi_{\text{n}}$ where, the first two terms denote that $\vtr{n}$ $ (\vtr{m_{\text{net}}})$ tends to be perpendicular (parallel) to $\vtr{H_{\text{ext}}}$. It results in an FLT as discussed in Section \ref{Six-state}. Upon simplification, $E_{\mathrm{total}} = H_{\mathrm{ext}}^{2}\cos^{2}(\phi_{\text{n}}-\phi_{\text{H}}) + m_{\mathrm{net}} H_{\mathrm{ext}}\sin(\phi_{\text{n}}-\phi_{\text{H}}) - K_{\text{ani}}\cos6\phi_{\text{n}}$. Now,  we assume that the triaxial anisotropy dominates Zeeman energy such that $K_{\text{ani}} \gg H_{\mathrm{ext}}^{2}, m_{\mathrm{net}} H_{\mathrm{ext}}$ and  $\vtr{n}$ stays close to one of the six energy minima, $\phi_{\text{0}}$. The easy directions $\phi_{\text{0}}$ are given by $n \pi/3$ where, $n = 0, 1, .., 5$. We consider a small deviation, $\delta\ll$ 1 from the easy axis such that $\phi_{\text{n}} = \phi_{\text{0}} + \delta$. Under the small angle $\delta$ expansion near the minimum, $E_{\text{total}}$ simplifies to, $E_{\mathrm{total}} = H_{\mathrm{ext}}^{2}(\cos^{2}(\phi_{\text{0}}-\phi_{\text{H}}) - \delta\sin2( \phi_{\text{0}}-\phi_{\text{H}})) + m_{\mathrm{net}} H_{\mathrm{ext}}(\sin(\phi_{\text{0}}-\phi_{\text{H}})+\delta\cos(\phi_{\text{0}}-\phi_{\text{H}}) ) - K_{\text{ani}} (1-18 \delta^{2})$. Minimizing the $E_{\mathrm{total}}(\delta)$ with respect to $\delta$ yields, $\phi_{\text{n}} \approx \phi_{\text{0}} + \frac{ H_{\text{ext}}^2 \sin 2 (\phi_{\text{0}}-\phi_{\text{H}}) - m_{\text{net}} H_{\text{ext}} \cos(\phi_{\text{0}}-\phi_{\text{H}})}{36K_{\text{ani}}}$. In the weak field limit ($H_{{ext}}^2\ll m_{\text{net}} H_{\text{ext}}$ i.e., $H_{\text{ext}}<m_{\text{net}}$), $\phi_{\text{n}}$ can be further reduced to
\begin{equation}
\label{eq:phin}
\phi_{\text{n}} \approx \phi_{\text{0}} + \frac{ - m_{\text{net}} H_{\text{ext}} \cos(\phi_{\text{0}}-\phi_{\text{H}})}{36K_{\text{ani}}}
\end{equation}
where, $\phi_{\text{n}}$ scales linearly with $H_{\text{ext}}$ about the value of $\phi_{\text{0}}$. For $\delta \ll 1$, $\rho_{xy}^{\mathrm{SMR}} \propto \frac{1}{2}\sin2\phi_{\text{n}}$ can be written as $\rho_{\text{xy}}^{\text{SMR}} \propto \frac{1}{2} \sin2\phi_{\text{0}} +\delta \cos 2\phi_{\text{0}}$. Further, substitution of $\phi_{\text{n}}$ (Eq. \ref{eq:phin}) in $\rho_{\text{xy}}^{\text{SMR}}$ expression modifies it to $\frac{1}{2} \sin2\phi_{\text{0}} - (\cos 2\phi_{\text{0}}) \big(\frac{m_{\text{net}}H_{\text{ext}}\cos(\phi_{\text{0}}-\phi_{\text{H}}) }{36K_{\text{ani}}}\big)$. From this expression, it follows that, $\rho_{xy}^{\,SMR} \propto - \cos 2\phi_{\text{0}} \cos(\phi_{\text{0}}-\phi_{\text{H}})$. Notably, in our experiments, all states exhibit similar sinusoidal $\rho_{\text{xy}}^{\text{SMR}}$ angular dependence at $H_{\text{ext}}$ = 250 Oe (Fig. \ref{fig:2}(d)), consistent with the weak-field regime.

\subsection{Dual-modulation SMR for thermal drift mitigation}\label{sideband}

\begin{figure}
\includegraphics[width=0.5\textwidth]{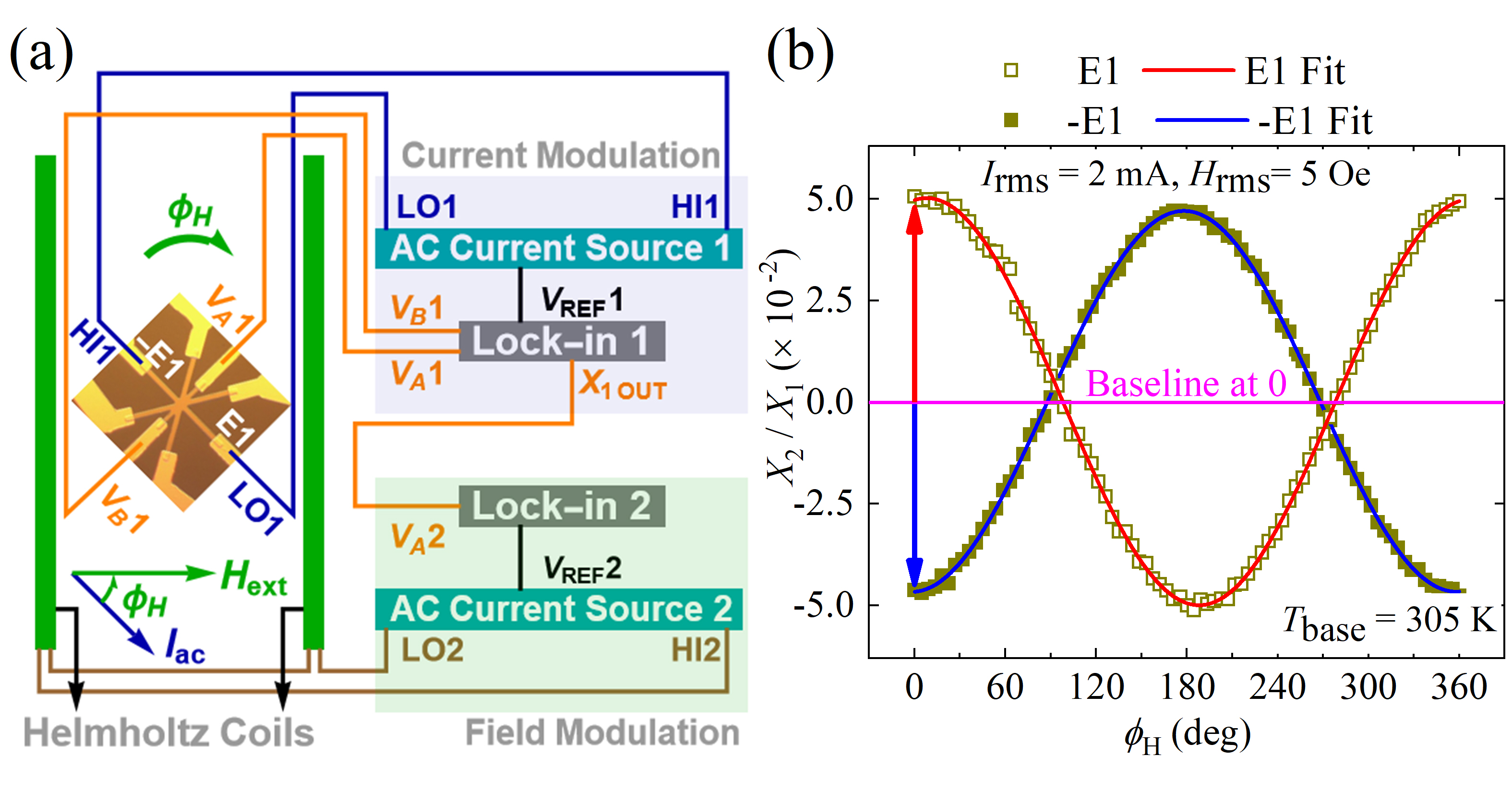}
\caption{\label{fig:3}(a) Schematic for simultaneous current and field modulation measurement.  (b) Dual-modulation SMR measurement  ($X_\text{2}$/$X_\text{1}$) distinguishes between \textbf{E1} and \textbf{-E1} states (maximum splitting marked at $\phi_{\text{H}}=0$ deg). Baseline of the $X_\text{2}$/$X_\text{1}$ modulation lies at zero and is temperature-independent. }
\end{figure}

Experimentally, we find that the SMR signal is highly sensitive to temporal temperature variations, hindering deterministic six-state probing. We analyze the source of this drift (see Appendix \ref{Thermaldrift}) and propose a dual-modulation (current and magnetic field) SMR method using two cascaded lock-in detections to distinguish the six states. In addition to the current-modulation ($I_\mathrm{ac} = \sqrt{2}I_{\mathrm{rms}}\sin(2\pi f_\text{I} t)$, $f_\text{I}=333$ Hz, device parameters: $l = 1.8 \times 10^{-4}$ m, $A = 2.3 \times 10^{-14}$ $\text{m}^2$), we simultaneously apply a second modulation, a magnetic field oriented at an angle $\phi_{\text{H}}$ in the $\textit{XY}$-plane, whose magnitude is modulated sinusoidally as $\vtr{H_{\text{ext}}}(t) =\sqrt{2} H_{\text{rms}} \sin(2\pi f_{\text{H}} t) (\cos\phi_{\text{H}} \hat{x} + \sin\phi_{\text{H}} \hat{y})$, $f_\text{H}=7$ Hz and $H_{\text{rms}} = 5$ Oe  (see Fig.\ref{fig:3}(a)). We used a custom-made Helmholtz coil and a Keithley 6221 AC current source for driving AC-fields. In weak-field regime, this modifies Eq. \ref{eq:phin} as $\phi_{\text{n}} (t) \approx \phi_{\text{0}} + \frac{- m_{\text{net}} H_{\mathrm{rms}} \sin(2\pi f_H t) \cos(\phi_{\text{0}}-\phi_{\text{H}}) }{36K_{\text{ani}}}$. Using $\phi_{\text{n}} (t)$ in $T$-dependent SMR form (Eq. \ref{eq:RhoxySMR}), we get

\begin{equation}
\label{eq:V(t)_compact}
\begin{split}
V_{\text{xy}}^{\mathrm{signal}}(t) = \sqrt{2}I_{\mathrm{rms}} \sin(2\pi f_\text{I} t)(l/A)\Big[
A_0\big(1 + \beta \Delta T(t)\big)+ A_H\\\big(1 + \beta \Delta T(t)\big)
\sin(2\pi f_\text{H} t)+ B_0\big(1 + \alpha_{\mathrm{Pt}} \Delta T(t)\big)
\Big]
\end{split}
\end{equation}
where,  $A_\text{H} = \frac{\sqrt{2}\Delta\rho\ m_{\text{net}} H_{\mathrm{rms}}}{36K_{\text{ani}}} \cos2\phi_{\text{0}}\cos(\phi_{\text{0}} - \phi_{\text{H}})$, $A_0 = -\frac{\Delta\rho}{2}\sin2\phi_{\text{0}}$, $B_0 = \epsilon \rho_0$, $\alpha_{Pt}$ and $\beta$ are the temperature coefficients of resistivity for \ch{Pt} and of SMR-amplitude, respectively. Expanding and rearranging terms,
\begin{equation}
\label{eq:V(t)_alternate}
\begin{split}
V_{\text{xy}}^{\mathrm{signal}}(t) =\sqrt{2}I_{\mathrm{rms}} \sin(2\pi f_\text{I} t)(l/A)\Big[
\rho_\text{0} + \rho_\text{T}(t) + \\ \rho_\text{H} \sin(2\pi f_\text{H} t) + \rho_\text{HT}(t)\sin(2\pi f_\text{H} t)
\Big]
\end{split}
\end{equation}
Here, $\rho_0 = A_0 + B_0$, $\rho_{\text{H}} = A_{\text{H}}$, $\rho_{\text{T}}(t) = (A_{\text{0}} \beta + B_{\text{0}} \alpha_{\text{Pt}}) \Delta T(t)$, and $\rho_{\text{HT}}(t) = A_{\text{H}} \beta \Delta T(t)$. Only the terms $\rho_{\text{T}}(t)$ and $\rho_{\text{HT}}(t)$ are sensitive to a change in $T$. The dual-modulated signal is fed into cascaded lock-in amplifiers, performing sequential demodulation to extract the $f_\text{H}$-dependent modulation of the $f_\text{I}$ response. The first lock-in output ($X_\text{1}(t)$) fed to the second lock-in can be written as $X_\text{1}(t) = \frac{I_{\mathrm{rms}}}{2}\frac{l}{A}\Big[
\rho_\text{0} + (A_0 \beta + B_0 \alpha_{\mathrm{Pt}})\Delta T_0
+ \rho_\text{H} (1 + \beta \Delta T_0)\sin(2\pi f_\text{H} t)\Big]$.

The measurable DC component in the first lock-in is $X_\text{1} =\frac{I_{\mathrm{rms}}}{2}\frac{l}{A}\Big[\rho_\text{0} + (A_\text{0} \beta + B_\text{0} \alpha_{\mathrm{Pt}})\Delta T_0\Big]$, while the second lock-in output is $X_\text{2} =\frac{I_{\mathrm{rms}}}{4}\frac{l}{A}\rho_\text{H} (1 +  \beta \Delta T_\text{0})$. Substituting all values in the ratio $X_\text{2} / X_\text{1}$ and normalizing by $\Delta \rho$ yields,
\begin{equation}
\label{eq:X1(t)}
\begin{split}
\frac{X_\text{2}}{ X_\text{1}} =\frac{
\dfrac{m_{\text{net}} H_{\mathrm{rms}}}{72 K_{\mathrm{ani}}}
\cos2\phi_{\text{0}}\cos(\phi_{\text{0}} - \phi_{\text{H}})\,(1+\beta \Delta T_0)
}{
\dfrac{-\sin2\phi_{\text{0}}}{2}
+ \dfrac{\epsilon \rho_0}{\Delta \rho}
+ \left(\dfrac{-\beta \sin2\phi_{\text{0}}}{2}
+ \dfrac{\epsilon \rho_0 \alpha_{\mathrm{Pt}}}{\Delta \rho}\right)\Delta T_0
}
\end{split}
\end{equation}
In the experimental limit, $\epsilon \rho_0 \ll \Delta \rho$ and $\beta \Delta T_0 \ll 1$ (weak heating/small $T$ change) yield the approximation,
\begin{equation}
\label{eq:X2/X1_final}
\frac{X_\text{2}}{X_\text{1}} \approx 
\frac{m_{\text{net}} H_{\mathrm{rms}}}{36 K_{\mathrm{ani}}}
\frac{\cos2\phi_{\text{0}}\,\cos(\phi_{\text{0}} - \phi_{\text{H}})}{\sin2\phi_{\text{0}}}
\end{equation}

Although the expression appears to diverge at $\phi_{\text{0}}=$ 0 and $\pi$, small thermal effects neglected in Eq. \ref{eq:X1(t)} render a finite response.  $X_\text{2}/X_\text{1}$ for the opposing states \textbf{E1} and \textbf{-E1} was extracted from the X-channels of two simultaneously measured lock-in amplifiers (Fig. \ref{fig:3}(b)). The signal exhibits sinusoidal variation with state-dependent phase shifts, consistent with Eq. \ref{eq:X2/X1_final}. The arrows in the Fig. \ref{fig:3}(b) show the maximum splitting of the $X_\text{2}/X_\text{1}$ signal for  \textbf{E1} and \textbf{-E1} states at 0 degrees. The linear dependence of $X_\text{2}/X_\text{1}$ with $H_{\text{rms}}$, consistent with Eq. \ref{eq:X2/X1_final}, is experimentally verified (Fig. S5 in \cite{supplemental}). Notably, the small amplitudes of $X_\text{2}/X_\text{1} \leq 0.05$ observed in the experiments are consistent with $\frac{m_{\text{net}} H_{\text{rms}}}{36 K_{\text{ani}}}\ll 1$. The baseline of $X_\text{2}/X_\text{1}$ signal remains at 0 and is invariant under different states. $X_\text{2}/X_\text{1}$ cancels out low frequency thermal drifts, making it a robust and reliable parameter for reading the states. We examined $X_\text{2}/X_\text{1}$ signals at 305 and 309 K ($\Delta T = 4$ K) and found that they coincide with each other demonstrating the effectiveness of the method in mitigating the thermal effects (refer Fig. S6 in \cite{supplemental}). Beyond SMR, the dual modulation approach is applicable to diverse measurements, including magneto-optic studies of SOT in \ch{YIG}$\vert$\ch{Pt} \cite{Montazeri2015}.

\subsection{Prospects of $2 \omega$ SMR for state readout}\label{2w}
\begin{figure}
\includegraphics[width=0.5\textwidth]{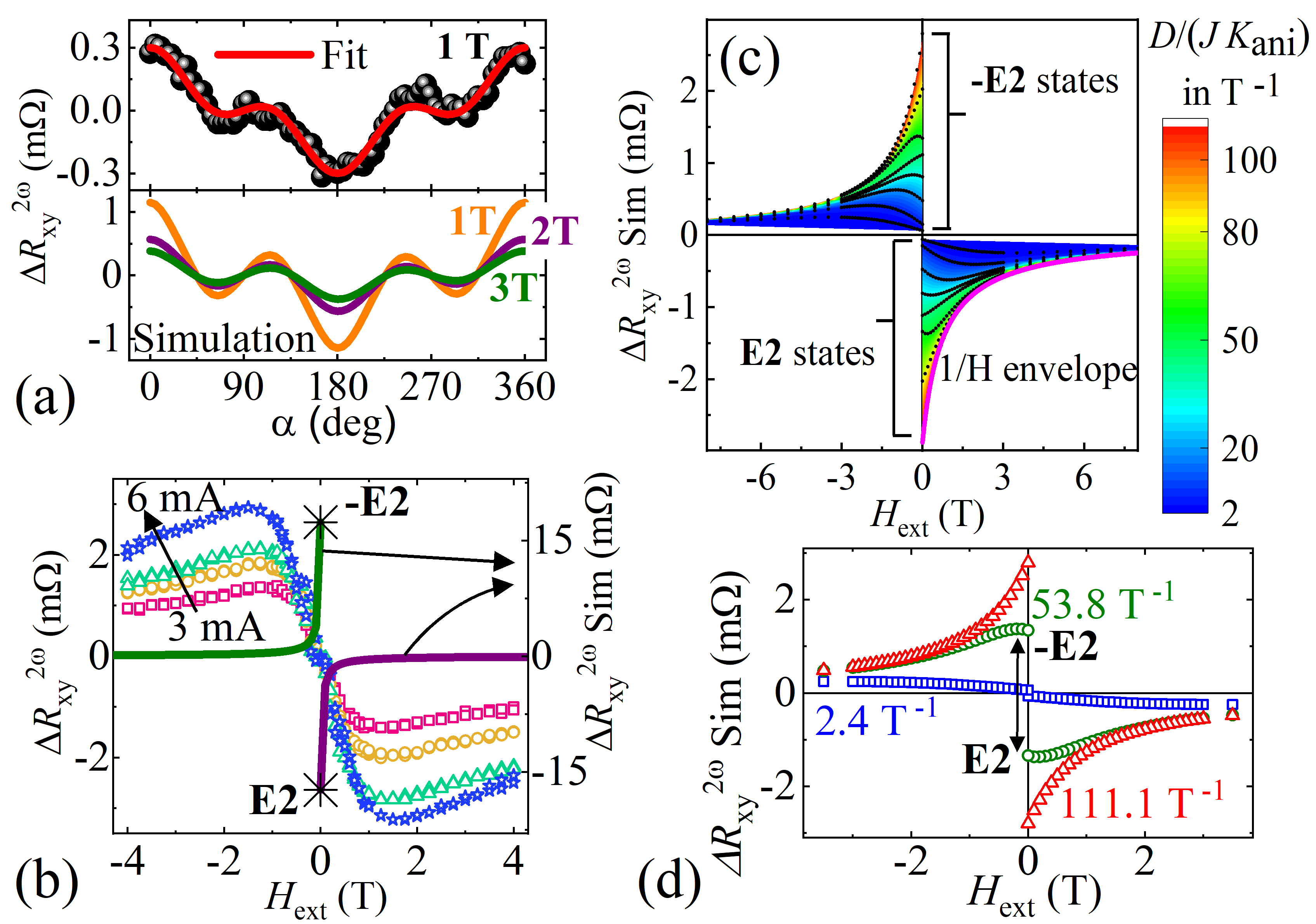}
\caption{\label{fig:4}(a) Top: $\alpha$-scan of 2$\omega$ SMR with fit considering field-like torque and spin Seebeck effect. Bottom: Simulation data for various $H_{\text{ext}}$ obtained using our model ($J$ = 850 T, $D$ = 2.5 T, and $K_{\text{plane}}$ = 1.7$\times 10^{-2}$ T \cite{Cogulu2022} along with $K_{\text{ani}} = 1$ $\mu\text{T}$). (b) Simulated $\Delta R_{\text{xy}}^{\,\text{2}\omega} \text{Sim}$ (right axis) vs. $H_{\text{ext}}$ curves using the above parameters. Experimental $\Delta R_{\text{xy}}^{\,\text{2}\omega}$ (left axis) vs. $H_{\text{ext}}$ data for various $I_{\text{rms}}$. (c) Simulated data sets of ${R_{\text{xy}}}^{2\omega}$ Sim vs. $H_{\text{ext}}$ for various $D/(J K_{\text{ani}})$ ratios. (d) Zoomed-in part of (c): simulations near $H_{\text{ext}} =$ 0 for selected three  $D/(J K_{\text{ani}})$ ratios.}
\end{figure}

The second harmonic (2$\omega$) SMR signal is influenced by the torque dynamics of the sublattice magnetization and is typically robust against thermal drifts. Here, we examine its potential for the reliable probing of six-state memory. First, we measured the angular dependence of 2$\omega$ SMR by in-plane rotation of the field $H_\text{ext}\gg H_{\text{ani}}$ (Fig. \ref{fig:4}(a)). Earlier work \cite{Cogulu2022} identified FLT as the dominant mechanism in $\alpha$-\ch{Fe2O3} dynamics. We infer that it is due to the strong easy-plane anisotropy ($K_{\text{plane}}$) and our experiments indicate a similar $\Delta R_{\text{xy}}^{\,\text{2}\omega}$ dependence on $\alpha$ at $H_{\text{ext}}$ = 1 T. The $\Delta R_{\text{xy}}^{\,\text{2}\omega}$ signal comprises of FLT ($\propto \cos2\alpha \cos\alpha$) and Spin Seebeck (SSE) ($\propto \cos\alpha$) contributions \cite{Cogulu2022}. Their combined fit to the data is shown in Fig. \ref{fig:4}(a).
Furthermore, we computed the characteristic wiggle in $\Delta R_{\text{xy}}^{\,\text{2}\omega} \text{Sim}$ using parameters $J$, $D$ and $K_{\text{plane}}$ \cite{Cogulu2022}, with an additional $K_{\text{ani}}$ of 1 $\mu$T using Eq. \ref{eq:HeffLLG}. The simulated FLT-driven response for different fields is shown at the bottom of Fig. \ref{fig:4}(a). Notably, the periodicity of the $R_{\text{xy}}^{\,\text{2}\omega}$ vs. $\alpha$ curve for $H_{\text{ext}} \gg H_\text{ani}$ is not 180$^\circ$. This indicates two distinct values of the signal for oppositely oriented spin-configurations and raises the question of whether the distinction can even be observed for $H_{\text{ext}} < H_\text{ani}$.

Motivated by this, we simulate $\Delta R_{\text{xy}}^{\,\text{2}\omega} \text{Sim}$ vs. $H_{\text{ext}}$ response by first initializing a single-domain \textbf{E2} (\textbf{-E2}) state with $H_{\text{ext}} \gg H_\text{ani}$ and decreasing $H_{\text{ext}}$ to 0 (Fig. \ref{fig:4}(b), right axis). In high-field regime ($H_{\text{ext}} \gg H_\text{ani}$), we see a characteristic 1/$H$-dependence consistent with the earlier work \cite{Cogulu2022}, arising from competing effective fields $H_{\text{ext}}$ and $H_{\text{FL}}^{\,\text{current}}$ due to external field and electric current, respectively. While, in low-field regime ($H_{\text{ext}} < H_\text{ani}$), an additional influence on the signal is seen due to effective field of tri-axial anisotropy ($K_{\text{ani}}=$ 1 $\mu$T in our case). Notably, at zero field, remenant values for \textbf{E2} and \textbf{-E2} are distinct. Experimentally, $\Delta R_{\text{xy}}^{\,2\omega}$ deviates from this behavior at low $H_{\text{ext}}$ with no clear detectable spontaneous splitting between \textbf{E2} and \textbf{–E2} (Fig. \ref{fig:4}(b) left axis) for various $I_{\text{rms}}$ values. To elucidate the state-splitting, we analyze $\Delta R_{\text{xy}}^{\,2\omega} \text{Sim} (H_{\text{ext}})$ response (Fig. \ref{fig:4}(c)) from our model for \textbf{E2} and \textbf{-E2} by varying $J$, $D$, and $K_{\text{ani}}$ through the ratio $D/(JK_{\text{ani}})$ over the relevant ranges 850--1050 T, 0.5--2.5 T, and 1--80 $\mu$T, respectively \cite{Yang2024, Cogulu2022, Sulymenko2017, Sheng2025, Lebrun2020, kanj2023, Li2026}. Large $D/(JK_{\text{ani}})$ ratio leads to pronounced splitting between \textbf{E2} and \textbf{-E2} at $H_{\text{ext}}=$ 0 while preserving the 1/$H$ envelope at high fields. Reducing this ratio suppresses the splitting (Fig. \ref{fig:4}(d)), and at low values (e.g., 2.4 $\text{T}^{-1}$), the remenant states become nearly indistinguishable, consistent with our observations as in Fig. \ref{fig:4}(b).

\begin{figure}
\includegraphics[width=0.5\textwidth,keepaspectratio]{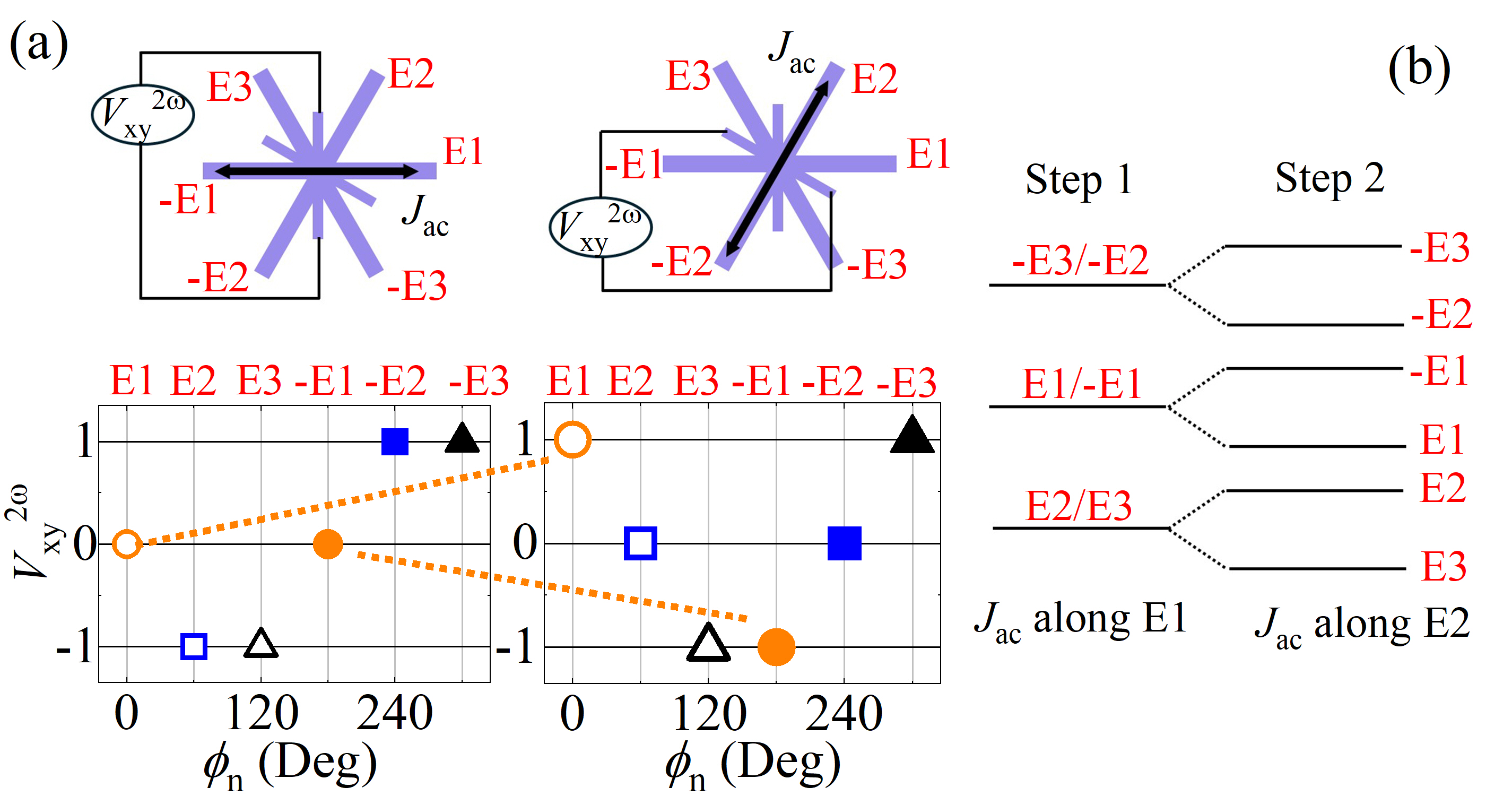}
\caption{\label{fig:5} Proposed two-step readout for six-state resolution in $\alpha$-\ch{Fe2O3}$\vert$\ch{Pt}. (a) Measurement geometry and corresponding $V_{\text{xy}}^{\,2\omega}$ for sequential state identification. Open circle, square and triangle represent \textbf{E1}, \textbf{E2} and \textbf{E3} states, respectively. The corresponding solid symbols denote the opposite states. b) Flow diagram illustrating degeneracy lifting via successive changes in current and voltage probe.}
\end{figure}
Hence, current-only distinguishability of \textbf{E} and \textbf{-E} states requires suitable tuning of the $D/(JK_{\text{ani}})$ ratio. Such tuning can be achieved via controlling thin-film deposition parameters \cite{Scheufele2023, Kan2022}, substrate-induced strain \cite{Harrison2025}, chemical doping \cite{Tanaka2024}, or thickness-dependent strain relaxation. Thus, we propose a current-only readout scheme for all six states in $\alpha$-\ch{Fe2O3} that involves two sequential steps (refer Fig. \ref{fig:5}): (1) $J_{\text{ac}}$ is applied along the \textbf{E1}/\textbf{-E1} axis, and $V_{\text{xy}}^{\,2\omega}$ signal is measured perpendicular to it; (2) $J_{\text{ac}}$ is then applied along the \textbf{E2}/\textbf{-E2} axis, and $V_{\text{xy}}^{\,2\omega}$ signal is measured with a second voltage-lead orthogonal to it. Numerical simulations for the protocol are carried out for a particular $J_{\text{ac}}$. In step one, the six memory states yield three distinct $V_{\text{xy}}^{\,2\omega}$ levels, mapped in $V_{\text{xy}}^{\,2\omega}$ vs. $\phi_{\text{n}}$ (Fig. \ref{fig:5}(a)). Step two lifts the remaining degeneracy, enabling the unambiguous identification of all states. This two-step protocol (Fig. \ref{fig:5}(b)) provides a deterministic, current-only readout and can be implemented via a logic tree or lookup-table architecture.

In summary, we demonstrate an unambiguous six-state readout in tri-axial canted AFM $\alpha$-\ch{Fe2O3} by measuring SMR under field-like perturbations ($H_\text{ext} \ll H_{\text{ani}}$). Under field-like perturbation, the spontaneous canting enables lifting of the degeneracy in SMR values of the opposite states, allowing their distinction. This is confirmed by $1\omega$ SMR and is further supported by numerical and analytical modeling by incorporating in-plane tri-axial anisotropy. Notably, temperature-induced drifts that limit reliable state distinguishability are effectively mitigated via dual-modulation SMR measurements employing combined current and field excitations. Additionally, our computations reveal that 2$\omega$ SMR is dependent on the coupled interplay between in-plane anisotropy, exchange and DM interactions and the ratio $D/(JK_{\text{ani}})$ governs the low-field and remanent response of 2$\omega$ SMR, providing a guideline for decisive degeneracy-lifting of opposite states. Finally, we propose a current-only two-step protocol for six-state readout using 2$\omega$ SMR.

\section*{ACKNOWLEDGEMENTS}

The authors acknowledge the DST, India, for providing funding support and are grateful to the National Facility for Low Temperature and High Magnetic Field, for facilitating some of the spin transport measurements. The authors acknowledge the DST-FIST funded Rigaku SmartLab X-Ray Diffractometer, Central Instruments Facility, Department of Physics, IISc, Bangalore. We acknowledge NNFC, CeNSE, IISc, Bangalore, supported by Government of India for lithography fabrication facility.

A.A.W. and S.G.B. contributed equally to this work.

\bibliography{fe2o3}

\setcounter{equation}{0}
\setcounter{figure}{0}
\renewcommand{\theequation}{A\arabic{equation}}
\renewcommand{\thefigure}{A\arabic{figure}}

\onecolumngrid
\begin{center}
\Large\bfseries End Matter
\end{center}
\twocolumngrid

\appendix

\section{Torque-balance equation}\label{LLG}
We model the temporal evolution of the sublattice magnetizations in $\alpha$-\ch{Fe2O3} under the influence of external field-like perturbation using the atomistic Landau-Lifshitz-Gilbert (LLG) equation, $\frac{d\vtr{M}_\text{1,2}}{dt}= -\gamma \vtr{M}_{\text{1,2}} \times \vtr{H}_\text{eff (1,2)} + \alpha_{\text{damp}} \vtr{M}_\text{1,2} \times \frac{d\vtr{M}_\text{1,2}}{dt}$. The effective fields, $\vtr{H_\text{eff(1,2)}}$, include contributions from exchange interaction ($\vtr{H_{\text{exch(1,2)}}}$), DM interactions ($\vtr{H_{\text{DM(1,2)}}}$), easy-plane anisotropy ($\vtr{H_{\text{plane(1,2)}}}$), six-fold in plane anisotropy ($\vtr{H_{\text{ani(1,2)}}}$) balancing the magnetization internally, and the perturbative external effective fields due to magnetic field ($\vtr{H_{\text{ext}}}$) and electric current ($\vtr{H_{\text{FL}}^{\text{current}}}$). 
\begin{widetext}
\begin{equation}
\label{eq:HeffLLG}
\begin{aligned}
    \vtr{H}_{\text{eff (1,2)}} = - J \vtr{M}_{2,1} \pm \vtr{D} \times \vtr{M}_{2,1} - K_{\text{plane}} \left(0, 0, {m}_{\text{(1,2)z}}\right)
    -6 K_{\text{ani}} \sin{6 \varphi_{1,2}} \left(\frac{-{m}_{\text{(1,2)y}}}{{{m}_{\text{(1,2)x}}}^{2}+{{m}_{\text{(1,2)y}}}^{2}}, \frac{{m}_{\text{(1,2)x}}}{{{m}_{\text{(1,2)x}}}^{2}+{{m}_{\text{(1,2)y}}}^{2}}, 0\right)\\
    + {H}_{\text{ext}}(\sin{{\theta}}\cos{{\alpha}},\sin{{\theta}}\sin{{\alpha}}, \cos{\theta}) + {h}_\text{FL} ({J}_{\text{ac}} \sin{2\pi ft}) \vtr{\sigma}
\end{aligned}
\end{equation}
\end{widetext}
where, sublattice magnetizations $\vtr{M_{1,2}}$ align antiparallel to each other with a small canting angle ${\delta}_{\text{Cant}}$ due to a combined effect of exchange interaction ($J$) and DM interaction ($\vtr{D}$, oriented along $\hat{z}$). Furthermore, we incorporate the easy-plane anisotropy (${K}_{\text{plane}}$) within the \textit{XY} plane. Notably, we add an effective field term corresponding to tri-axial anisotropy within the \textit{XY} plane, characterized by an anisotropy constant, ${K}_{\text{ani}}$. The formulation of $\vtr{H}_{\text{ani}}$ used in Eq. \ref{eq:HeffLLG} is detailed in \cite{supplemental}. The angle ${\varphi}_{1,2}$ used to estimate $\vtr{H}_{\text{ani}}$, represents the angle between $\vtr{M}_{1,2}$ and $\hat{x}$. Further, the last two terms in Eq. \ref{eq:HeffLLG}, represent effective fields, $\vtr{H}_{\text{ext}}$ and $\vtr{H}_{\text{FL}}^\text{current}$ respectively, originating from the external perturbations to the system. The reference angle of the applied magnetic field, ${\theta}$, is defined from $\hat{z}$ relative to the \textit{XY} plane. Further, $\vtr{H}_{\text{FL}}^{\text{current}}$ is time-dependent, oscillating at the frequency $f$ of the applied current density ${J}_{\text{ac}}$. The direction of this $\vtr{H}_{\text{FL}}^{\text{current}}$ is along the spin-accumulation vector near the interface $\vtr{\sigma}$, oriented along $\hat{y}$, primarily arising from the FLT term ($\vtr{M}_{1,2} \times \vtr{\sigma}$). The damping-like torque term ($\vtr{M}_{1,2} \times(\vtr{M}_{1,2} \times \vtr{\sigma})$) has a negligible impact on the solution to our LLG equation, consistent with earlier report \cite{Cogulu2022} in the \textit{XY} plane.

\setcounter{equation}{0}
\setcounter{figure}{0}
\renewcommand{\theequation}{B\arabic{equation}}
\renewcommand{\thefigure}{B\arabic{figure}}

\section{Thermal drift in SMR}\label{Thermaldrift}
For $\alpha$-\ch{Fe2O3}$\vert$\ch{Pt} with in-plane anisotropy in the presence of the truly in-plane field, the transverse SMR can be written as $\rho_{\text{xy}}^{\mathrm{SMR}} = -\Delta\rho \cos\alpha \sin\alpha$ \cite{Chen2013}. Due to asymmetry in the Hall-bar structure, a small portion of longitudinal SMR ($\rho_{\text{xx}}^{\mathrm{SMR}} = \rho_0 + \Delta\rho \cos^2\alpha$) can leak into $\rho_{\text{xy}}$ such that $\rho_{\text{xy}}^{\mathrm{meas}} = \rho_{\text{xy}}^{\mathrm{SMR}} + \epsilon \rho_{\text{xx}}^{\mathrm{SMR}}$ where $\epsilon \ll 1$. Substitution gives $\rho_{\text{xy}}^{\mathrm{meas}} = -\Delta\rho \cos\alpha \sin\alpha + \epsilon \rho_0 + \epsilon \Delta\rho \cos^2\alpha$ where, the last term is negligible. Taking into account the $T$-dependence of the resistivity of \ch{Pt} and of the SMR amplitude, it can be expressed in terms of the orientation of $\vtr{n}$ ($\phi_{\text{n}}$) as,
\begin{equation}
\label{eq:RhoxySMR}
\rho_{\text{xy}}^{\mathrm{meas}} = -\Delta \rho (1+\beta \Delta T) \cos \phi_{\text{n}} \sin \phi_{\text{n}} + \epsilon \rho_0 (1 + \alpha_{\text{Pt}} \Delta T)
\end{equation}
where, $\alpha_{\text{Pt}}$ and $\beta$ are the temperature coefficients of resistivity for \ch{Pt} and of SMR-amplitude, respectively. SMR amplitude varies with $T$ due to the change in spin Hall angle and spin-diffusion length (spin transport in metal) and spin-mixing conductance (interfacial spin transport).

\end{document}



\title{Supplemental Material\\ Field-like Perturbation Enabled Six-state Readout in Triaxial $\alpha$-\ch{Fe2O3}$\vert$\ch{Pt} Bi-layers}

\author{Aditya A. Wagh}
\affiliation{%
 Department of Physics, Indian Institute of Science, Bangalore, INDIA 
}%

\author{Shwetha G. Bhat}
\affiliation{%
Department of Physics, Indian Institute of Science, Bangalore, INDIA 
}%

\author{Krishna Jha}
\affiliation{%
Department of Physics, Indian Institute of Science, Bangalore, INDIA 
}%

\author{Aiswarya Sukumaran}
\affiliation{%
Department of Physics, Indian Institute of Science Education and Research, Tirupati, INDIA 
}%

\author{P. S. Anil Kumar}
\affiliation{%
Department of Physics, Indian Institute of Science, Bangalore, INDIA 
}%

\maketitle

\setcounter{equation}{0}
\renewcommand{\theequation}{S\arabic{equation}}

\setcounter{figure}{0}
\renewcommand{\thefigure}{S\arabic{figure}}

\setcounter{table}{0}
\renewcommand{\thetable}{S\arabic{table}}

\section{Pulsed Laser Deposition}
\label{PLD}
 $\alpha$-\ch{Fe2O3} $\vert$ Pt thin films are deposited on \ch{Al2O3} (0 0 0 1) substrate, using a home-grown target of $\gamma-$\ch{Fe2O3}. KrF excimer laser (248 nm) is used to ablate the target with an energy density $\approx$ 1.8 J/$\text{cm}^2$ deposition at a repetition rate of 3 Hz. The growth temperature of $\alpha$-\ch{Fe2O3} is maintained at 500 $^\circ$C at a pressure of \ch{O2} of 0.02 mbar. The films are annealed under the same conditions for 1 hour and then cooled to room temperature. A thin layer of Pt is later deposited at room temperature with a base pressure of 1 × $10^{-5}$ mbar. 
 \begin{figure}[hbt!]
\includegraphics[width=1.0\textwidth,keepaspectratio]{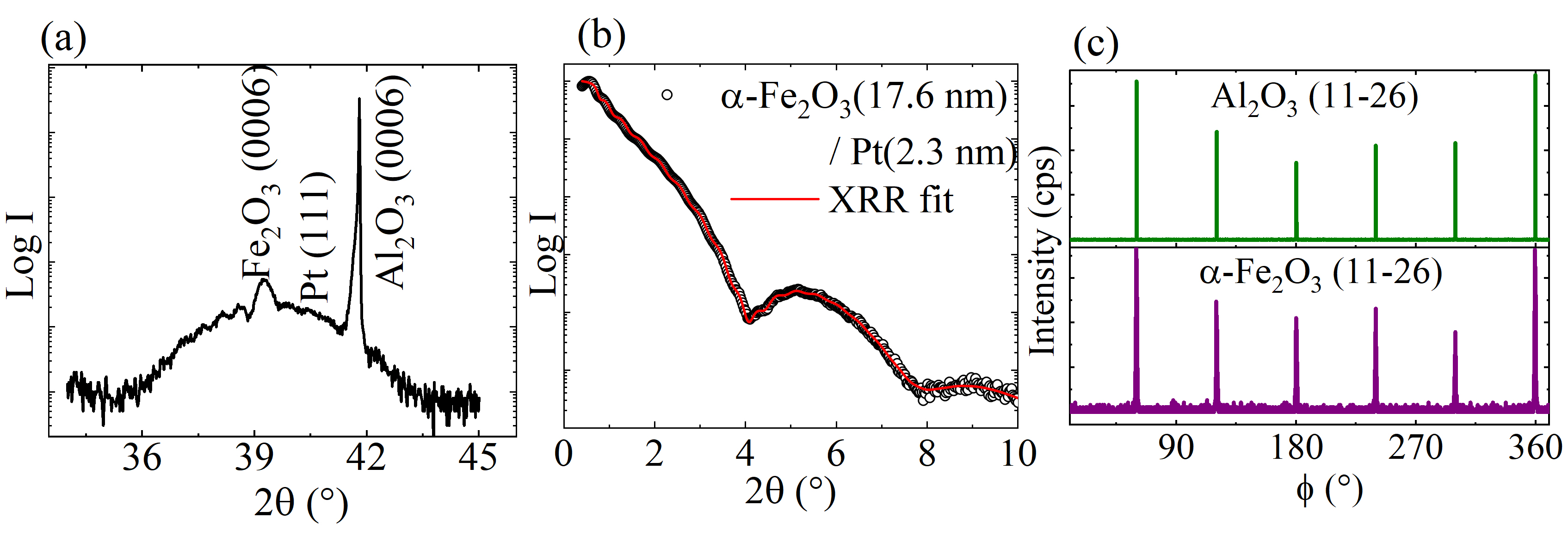}
\caption{\label{fig:XRD}(a) 2$\theta-\theta$ diffraction pattern of \ch{Al2O3} $\vert$ $\alpha$-\ch{Fe2O3} (17.6 nm) $\vert$ Pt (2.3 nm), (b) X-ray reflectivity curve for the thin film along with fitted curve. (c) The $\phi-$scan of the film exhibiting epitaxial growth of $\alpha$-\ch{Fe2O3} on \ch{Al2O3}.}
\end{figure}
 The structural characterization of the films was confirmed using X-ray diffraction (XRD), and Fig. \ref{fig:XRD}(a) exhibits the XRD pattern for (0 0 0 1) oriented $\alpha$-\ch{Fe2O3} (17.6 nm)|Pt (2.3 nm). The film thickness is confirmed using the X-ray reflectivity technique, and Fig. \ref{fig:XRD}(b) shows the film reflectivity data along with the fitted curve. Furthermore, the epitaxial growth of $\alpha$-\ch{Fe2O3} on \ch{Al2O3} is confirmed by performing the $\phi$ scans about (1 1 $\overline{2}$ 6) and the corresponding data are shown in Fig. \ref{fig:XRD}(c).

\section{Writing a state with magnetic field}
\label{Writing with H}
\begin{figure}[hbt!]
\includegraphics[width=1.05\textwidth,keepaspectratio]{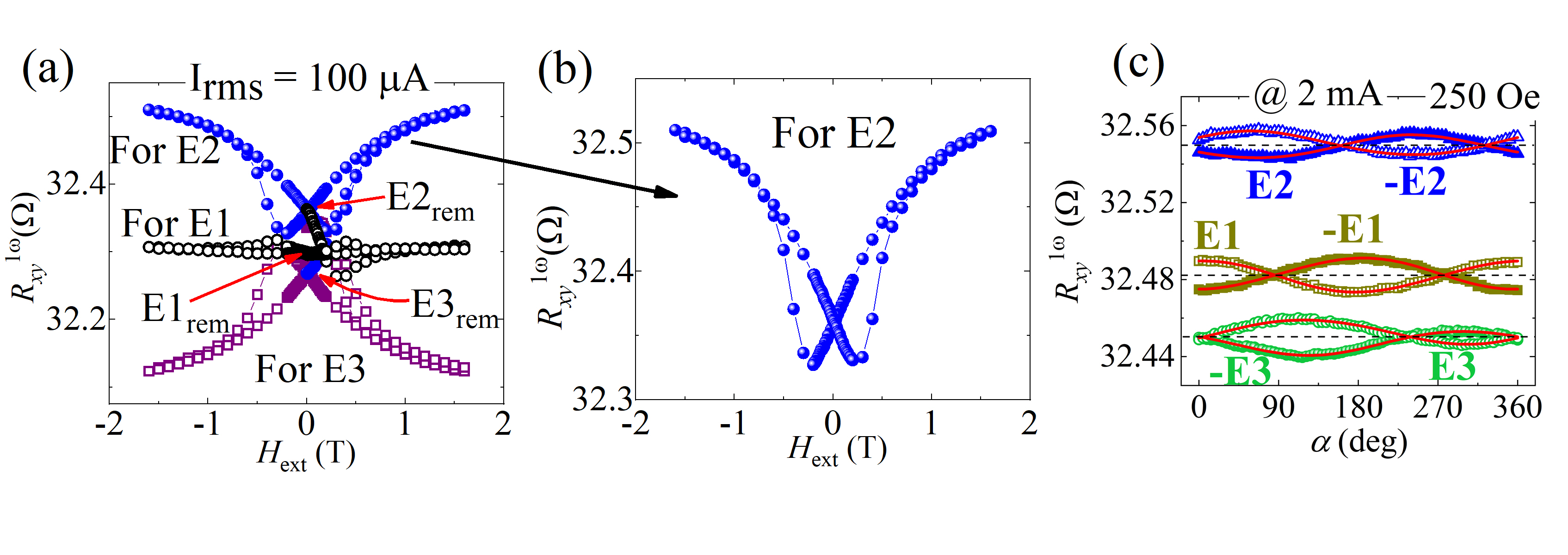}
\caption{\label{fig:1w} (a) $R_{\text{xy}}^{1\omega}$ vs. $H_{\text{ext}}$ corresponding to $\textbf{E1}$, $\textbf{E2}$, and $\textbf{E3}$ state measured for $I_{rms}$ = 100 $\mu$A. (b) $R_{\text{xy}}^{1\omega}$ vs. $H_{\text{ext}}$ for state $\textbf{E2}$ shown separately for clarity. (c) $R_{\text{xy}}^{1\omega}$ vs. $\alpha$ at $H_{\text{ext}}$ = 250 Oe and $I_{\text{rms}}$ = 2 mA for all six states.}
\end{figure}

The three distinct memory states ($\textbf{E1}$, $\textbf{E2}$, and $\textbf{E3}$) can be individually written in a $\alpha$-\ch{Fe2O3} Hall bar device with the help of a magnetic field, similar to the eight-terminal device presented in Section III of the paper. We measured first-harmonic (1$\omega$) transverse spin Hall magnetoresistance (SMR) ($R_{\text{xy}}^{1\omega}$) as a function of the applied magnetic field ($H_{\text{ext}}$ and the data is shown in Fig. \ref{fig:1w}(a). $H_{\text{ext}}$ is applied along $\vtr{m}_{\text{net}}$ and perpendicular to $\vtr{n}$ for each state. Fig. \ref{fig:1w}(b) is $R_{\text{xy}}^{1\omega}$ corresponding to the state $\textbf{E2}$ shown separately, for clarity. Additionally, Fig. \ref{fig:1w} (c) shows the angular response of $R_{\text{xy}}^{1\omega}$ at $H_{\text{ext}}$ = 250 Oe from our Hall device, which is consistent with the angular response we obtained from our eight-terminal device used in the main paper. 

\section{Six-fold Anisotropy in the basal plane of $\alpha$-F\MakeLowercase{\text{e}}$_2\text{O}_3$}
\label{6fold}
$\alpha$-\ch{Fe2O3} is known to associate with a six-fold anisotropy within its (0001) plane \cite{Cheng2020}. The anisotropy energy corresponding to the six-fold configuration in \textit{xy} plane is given by,

\begin{equation}
\label{eq:Eani}
\begin{aligned}
{E}_{\text{ani(1,2)}} = -{K}_{\text{ani(1,2)}} \cos(6\varphi)
\end{aligned}
\end{equation}

The effective field $H_{\text{ani}}$ due to the six-fold anisotropy can be defined as
\begin{subequations}
\label{eq:Hani}
\begin{equation}
 \vtr{H}_{\text{ani(1,2)}} = -\frac{\partial E_{\text{ani}}}{\partial  \vtr{M}_{1,2}}
= -6{K}_{\text{ani(1,2)}} \sin(6\varphi)\\    
\end{equation}
\begin{equation}
\vtr{H}_{\text{ani(1,2)}} = -6 K_{ani} \sin{6 \varphi_{1,2}} \left(\frac{-{m}_\text{(1,2)y}}{{{m}_\text{(1,2)x}}^{2}+{{m}_\text{(1,2)y}}^{2}}, \frac{{m}_\text{(1,2)x}}{{{m}_\text{(1,2)x}}^{2}+{{m}_\text{(1,2)y}}^{2}}, 0\right)
\end{equation}
\end{subequations}

\section{Simulation of angular dependence of $\Delta R_\text{xy}^{1\omega}$ for $H_\text{ext} \gg H_\text{ani}$}\label{sin2alpha}

\begin{figure}[hbt!]
\includegraphics[width=0.5\textwidth,keepaspectratio]{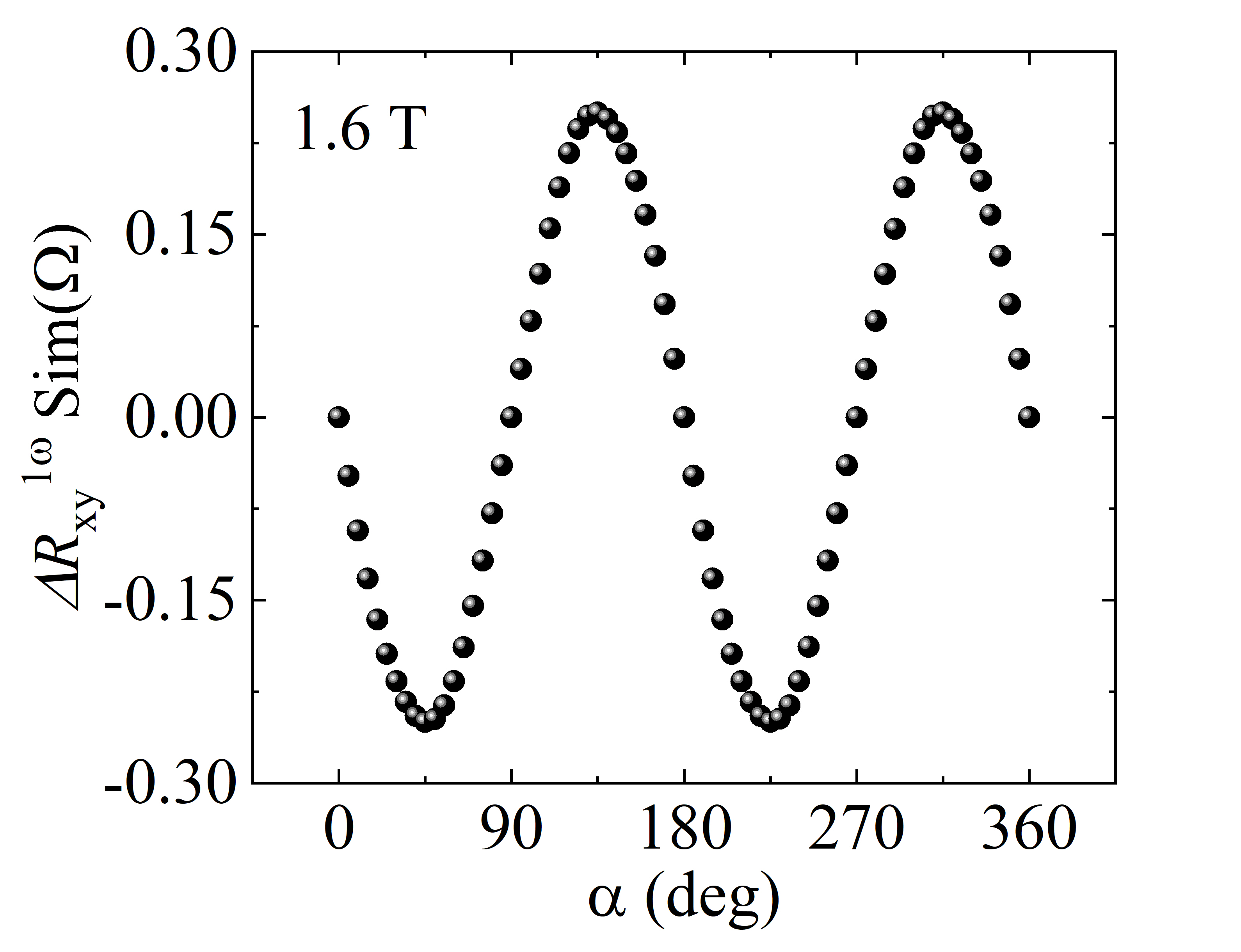}
\caption{\label{fig:sin2alpha} Simulated $\Delta R_\text{xy}^{1\omega}$ vs. $\alpha$ for $H_\text{ext}$ of 1.6 T and $I_\text{rms}$ of 2 mA.
}
\end{figure}

We simulated the angular dependence of $\Delta R_\text{xy}^{1\omega}$ for applied magnetic fields higher than the switching field to verify the general $\sin{2\alpha}$ behavior observed from our experiments. The simulated data is shown in Fig. \ref{fig:sin2alpha} for $H_\text{ext}$ of 1.6 T and $I_\text{ac}$ of 2 mA. This functional behavior is consistent with our experimental observation as discussed in the main paper.

\section{Simulation of $\Delta R_\text{xy}^{1\omega}$ signal for collinear antiferromagnet with tri-axial anisotropy}\label{collinear}

\begin{figure}[hbt!]
\includegraphics[width=0.6\textwidth,keepaspectratio]{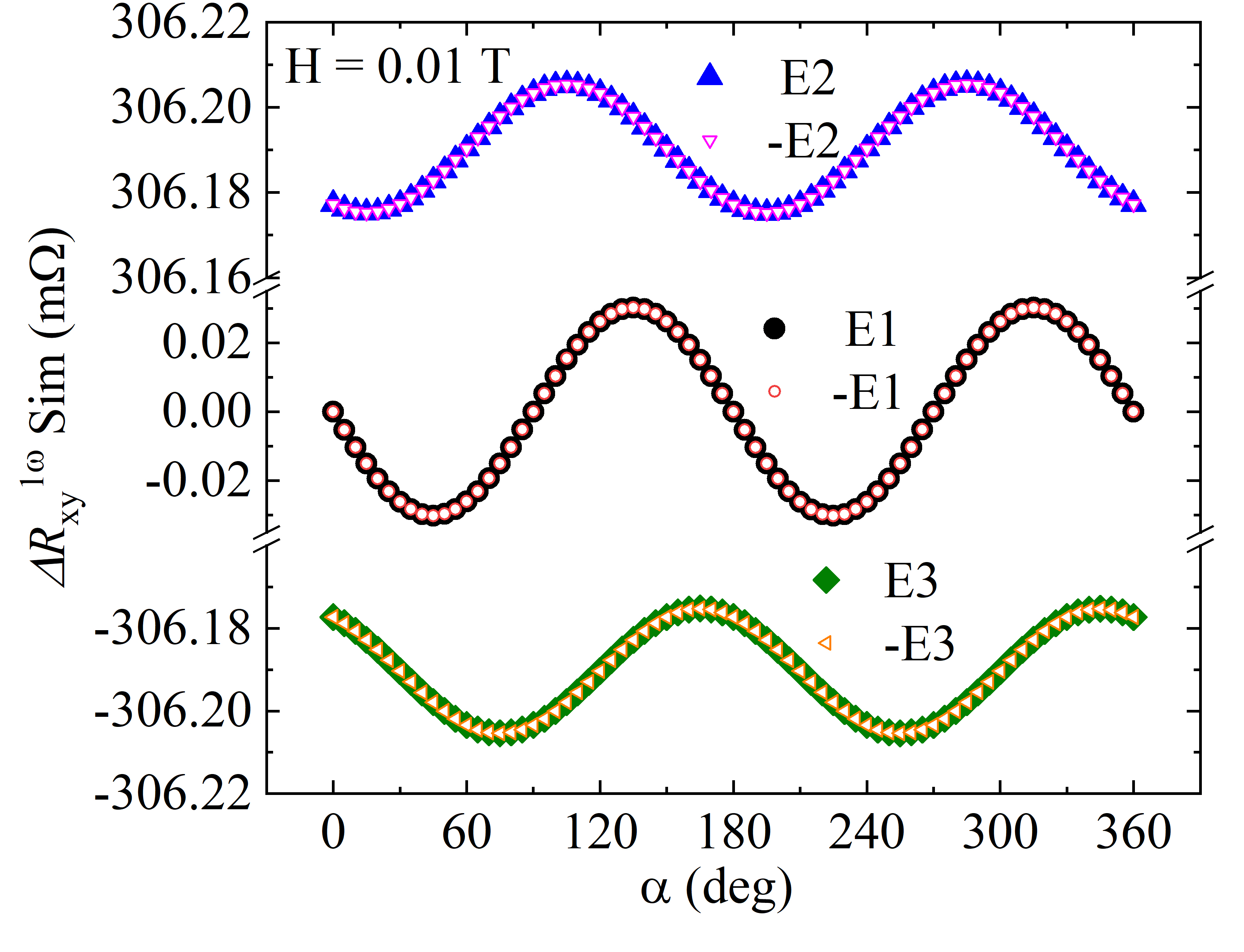}
\caption{\label{fig:collinear} Simulated angular scan of $\Delta R_\text{xy}^{1\omega}$ signal corresponding to collinear AFM in the absence of DM interaction for an applied field of 0.01 T.
}
\end{figure}

To validate our model, we extended the simulations to a collinear AFM system and compared the response with that of the canted AFM case. In the absence of DM interactions in collinear AFMs, we examined the angular dependence of the $\Delta R_\text{xy}^{1\omega}$ signal under applied fields below the critical value. The results shown in Fig. \ref{fig:collinear} display the corresponding response for all six degenerate states. Importantly, the SMR values for opposite states $\textbf{E}i$ and $\textbf{-E}i$ are identical, and the signal does not provide distinct signatures for the opposite states. This highlights that spontaneous canting, induced by DMI, is essential to eliminate degeneracy and allow for the distinction between all six states. Our findings thus underscore the crucial role of non-collinearity in accessing complete state resolution in AFM systems.

\section{Thermal drift mitigation using Dual Modulation SMR}

\begin{figure}[hbt!]
\includegraphics[width=0.7\textwidth,keepaspectratio]{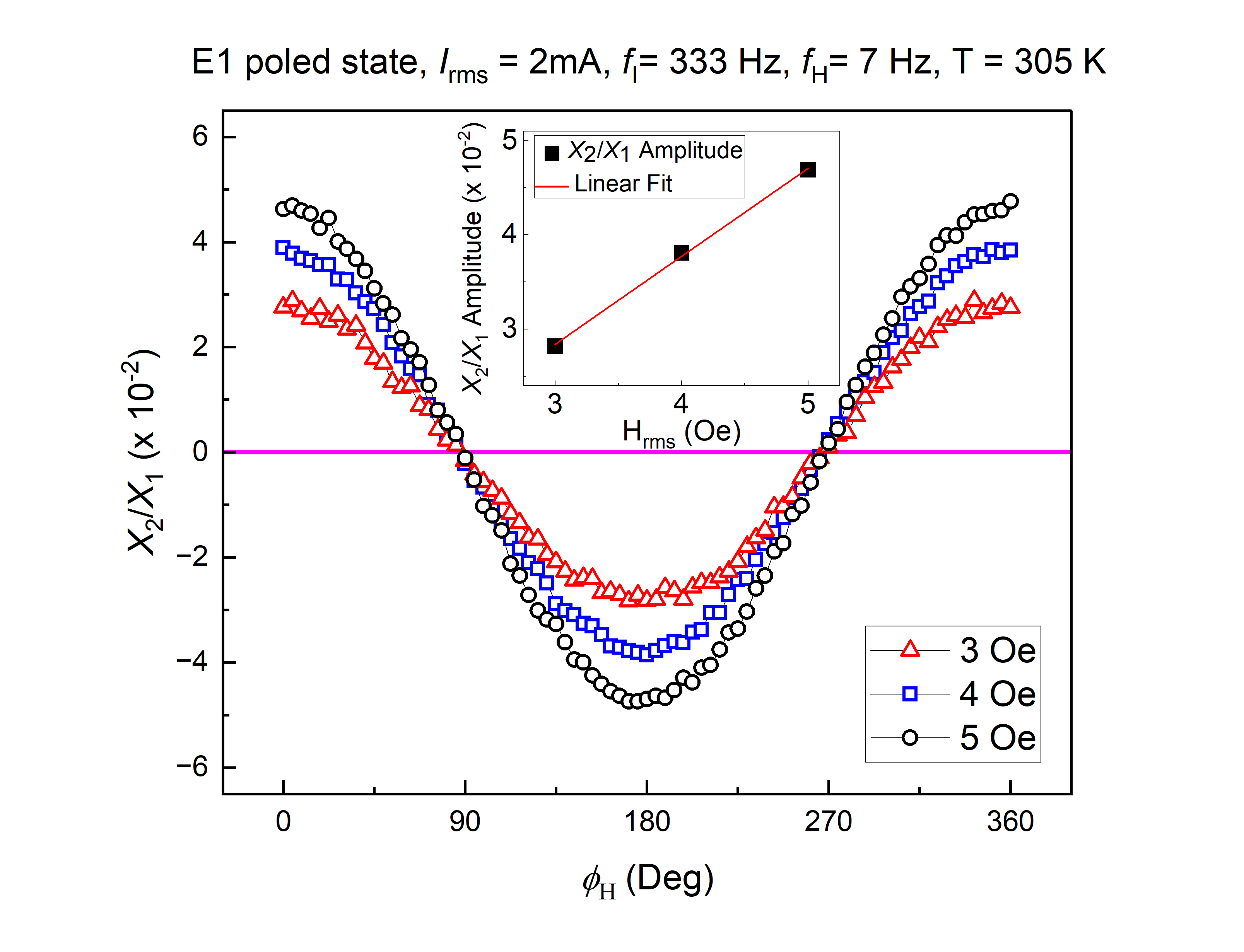}
\caption{\label{fig:Hdep} Angular scan of $X_2/X_1$ signal for different field values at T = 305 K.}
\end{figure}

We first measured dual modulation SMR ($X_2/X_1$) signal at three different field amplitudes ($H_{\text{rms}} = 3, 4 \text{ and } 5 \text{ Oe}$) (see Fig. \ref{fig:Hdep}). The inset of the Fig. \ref{fig:Hdep} shows that the amplitude of $X_2/X_1$ scales linearly with $H_{\text{rms}}$ as expected.

In this section, we examine and compare the influence of change in temperature on the SMR ($R_{\text{xy}}^{1\omega}$) and the dual modulation SMR ($X_2/X_1$) signals. The thermal drift of the $R_{\text{xy}}^{1\omega}$ signal can be described using the following equation (reproduced from Appendix B in the main text): 
\begin{equation}
\label{eq:TempdepSMR}
\begin{aligned}
\rho_{\text{xy}}^{\mathrm{meas}} = -\Delta \rho (1+\beta \Delta T) \cos \phi_{\text{n}} \sin \phi_{\text{n}} + \epsilon \rho_0 (1 + \alpha_{\text{Pt}} \Delta T)
\end{aligned}
\end{equation}

\begin{figure}[hbt!]
\includegraphics[width=1.0\textwidth,keepaspectratio]{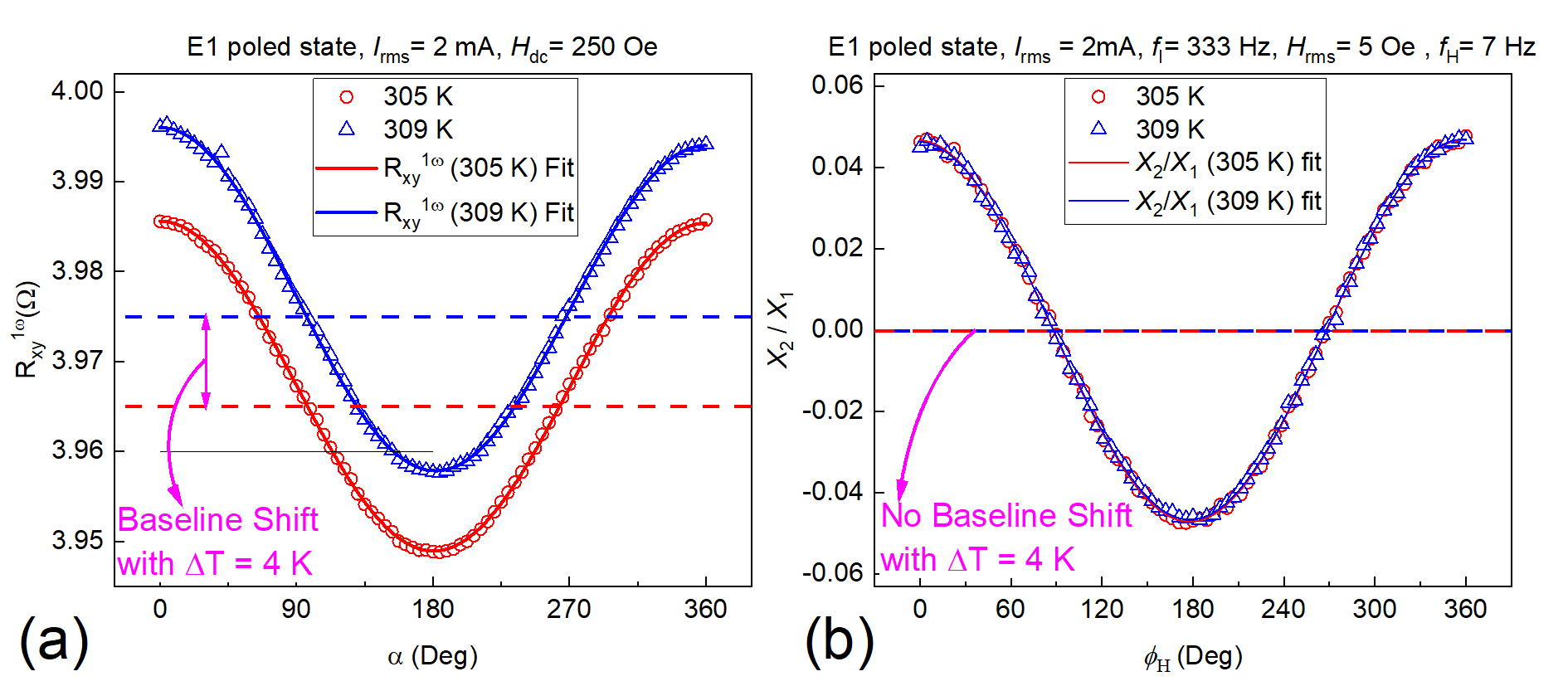}
\caption{\label{fig:tempdep} (a) Angular variation of $R_\text{xy}^{1\omega}$ at T = 305 K and 309 K. A clear shift in the baseline is marked by the arrow. (b) Angular scans of $X_2/X_1$ at the same two temperatures. $X_2/X_1$ signals and their baselines coincide with each other. 
}
\end{figure}

The large value of temperature coefficient of resistivity of \ch{Pt} ($\alpha_{\text{Pt}}$) makes the second term of the equation more sensitive to temperature changes (as $\epsilon\alpha_{Pt}\gg\beta\Delta\rho$). Slowly varying temperatures (very low-frequency thermal drifts) were observed to affect angular scans of $R_\text{xy}^{1\omega}$. In order to reject such low-frequency thermal drifts in the SMR signals, we employed dual modulation SMR method. In Fig. \ref{fig:tempdep} we present angular variation of  $R_\text{xy}^{1\omega}$ and $X_2/X_1$ measured at 305 and 309 K. A considerable change in temperature ($\Delta T = 4$ K) manifests as a discernible shift in the baseline of $R_\text{xy}^{1\omega}$ signal (see Fig. \ref{fig:tempdep}(a)). In contrast, the angular variations of $X_2/X_1$ measured at 305 and 309 K coincide with each other (see Fig. \ref{fig:tempdep}(b)) indicating that the dual modulation SMR method is effective in mitigating thermal drifts in the signal. 

\bibliography{fe2o3}